\documentstyle[12pt,twoside,fleqn,espcrc1p,epsfig,rotate,amssymb]{article}

\newcommand{\AmS}{{\protect\the\textfont2
  A\kern-.1667em\lower.5ex\hbox{M}\kern-.125emS}}

\def\preprint{ADP-00-41/T424}
\def\archive{nucl-th/0007058}

\title{Lattice QCD Calculations of Hadron Structure:\\
       Constituent Quarks and Chiral Symmetry}

\author{Derek B. Leinweber${}^{\rm \, a}$\thanks{Presented at Few Body
        2000, XVIth IUPAP International Conference on Few-Body
        Problems in Physics, National Taiwan University, Taipei,
        Taiwan, March 6--10, 2000.}
and 
Anthony W. Thomas\address{Special Research Center for the
Subatomic Structure of Matter (CSSM) and\\ Department of Physics and
Mathematical Physics, University of Adelaide 5005}
}
       
\begin{document}

\maketitle

\begin{abstract}
New data from parity-violating experiments on the deuteron now allow
isolation of the strange-quark contribution to the nucleon magnetic
moment, $G_M^s(0)$, without the uncertainty surrounding the anapole
moment of the nucleon.  Still, best estimates place $G_M^s(0) > 0$.
It is illustrated how this experimental result challenges the very
cornerstone of the constituent quark model.  The chiral physics giving
rise to $G_M^s(0) \sim 0$ is illustrated.
\end{abstract}

\section{INTRODUCTION}

Previous experimental reports \cite{SAMPLE} of the strange-quark
contribution to the nucleon magnetic moment, $G_M^s(0)$, have involved
the poorly understood nucleon anapole moment.  However, combining the
original proton data with new data from parity violation experiments
on the deuteron now allow an independent isolation of $G_M^s(0)$.  The
preliminary analysis \cite{Pitt} indicates $G_M^s(0)$ is smaller than
the earlier result \cite{SAMPLE} of
\begin{equation}
G_M^s(0.1\ {\rm GeV}^2) = +0.61 \pm 0.17 \pm 0.21 \mu_N \, ,
\end{equation}
but positive values for $G_M^s(0)$ are still favoured.

In this report, we elucidate the impact of this experimental result on
our understanding of the concept of a constituent quark.  In
particular, the constituent quark approach places properties such as
the magnetic moment contribution of a quark to a baryon intrinsic to
the constituent quark itself and therefore not dependent upon the
environment in which the quark resides.

The extraction of $G_M^s(0)$ is based on the approximation of charge
symmetry in the nucleon moments; i.e.\ equal $u$ and $d$ current quark
masses and negligible electromagnetic corrections.  This approximation
is expected to be valid at the 1\% level \cite{ChargeSymm} and is
certainly sufficient for the analysis of $G_M^s(0)$.

\begin{figure}[tbp]
\vspace{-1.5cm}
\begin{center}
\setlength{\unitlength}{1.0cm}
\setlength{\fboxsep}{0cm}
\begin{picture}(10,5)
\put(0,0){\begin{picture}(5,5)\put(0,0){
\rotate{\epsfig{file=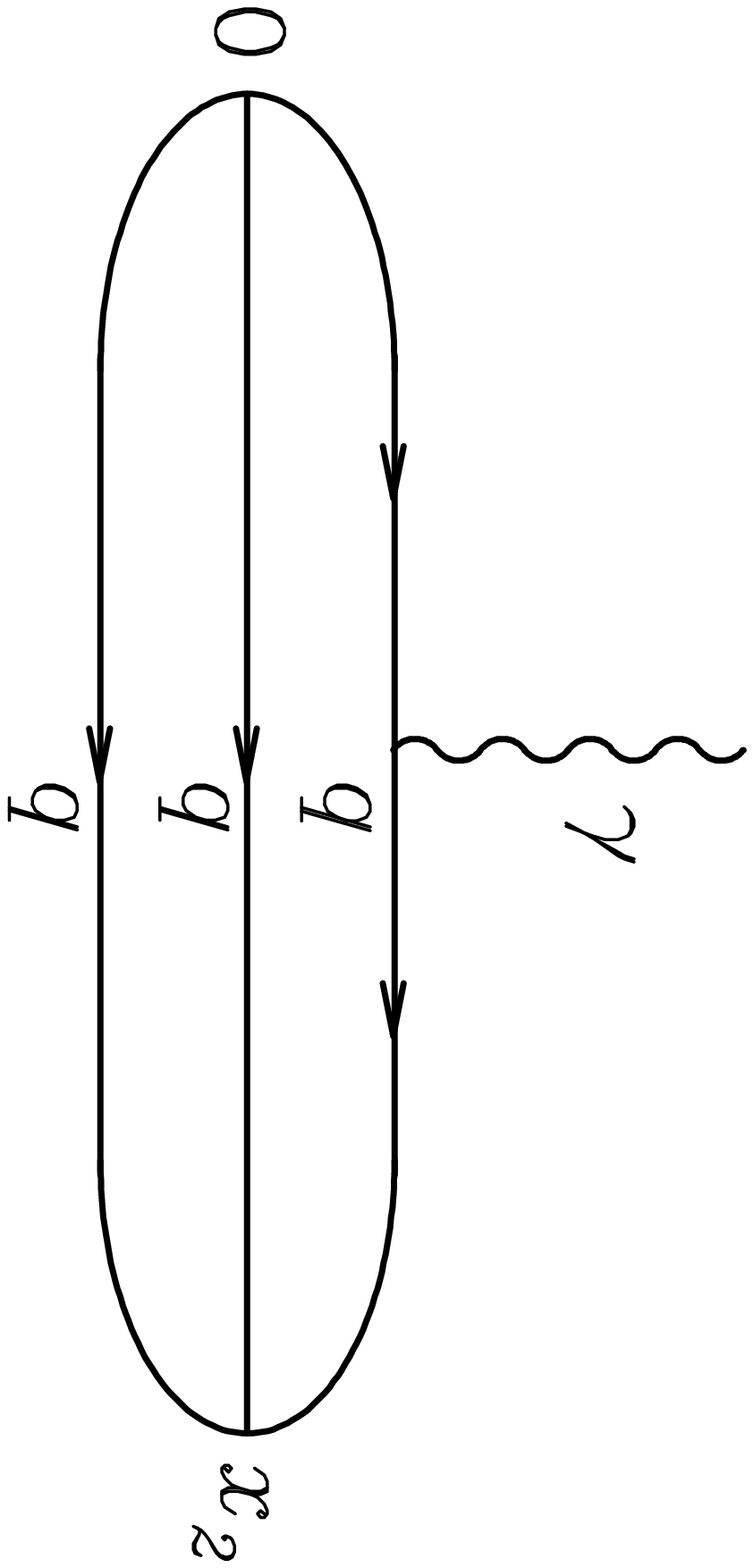,height=4cm}}}\end{picture}}
\put(5.5,0){\begin{picture}(5,5)\put(0,0){
\rotate{\epsfig{file=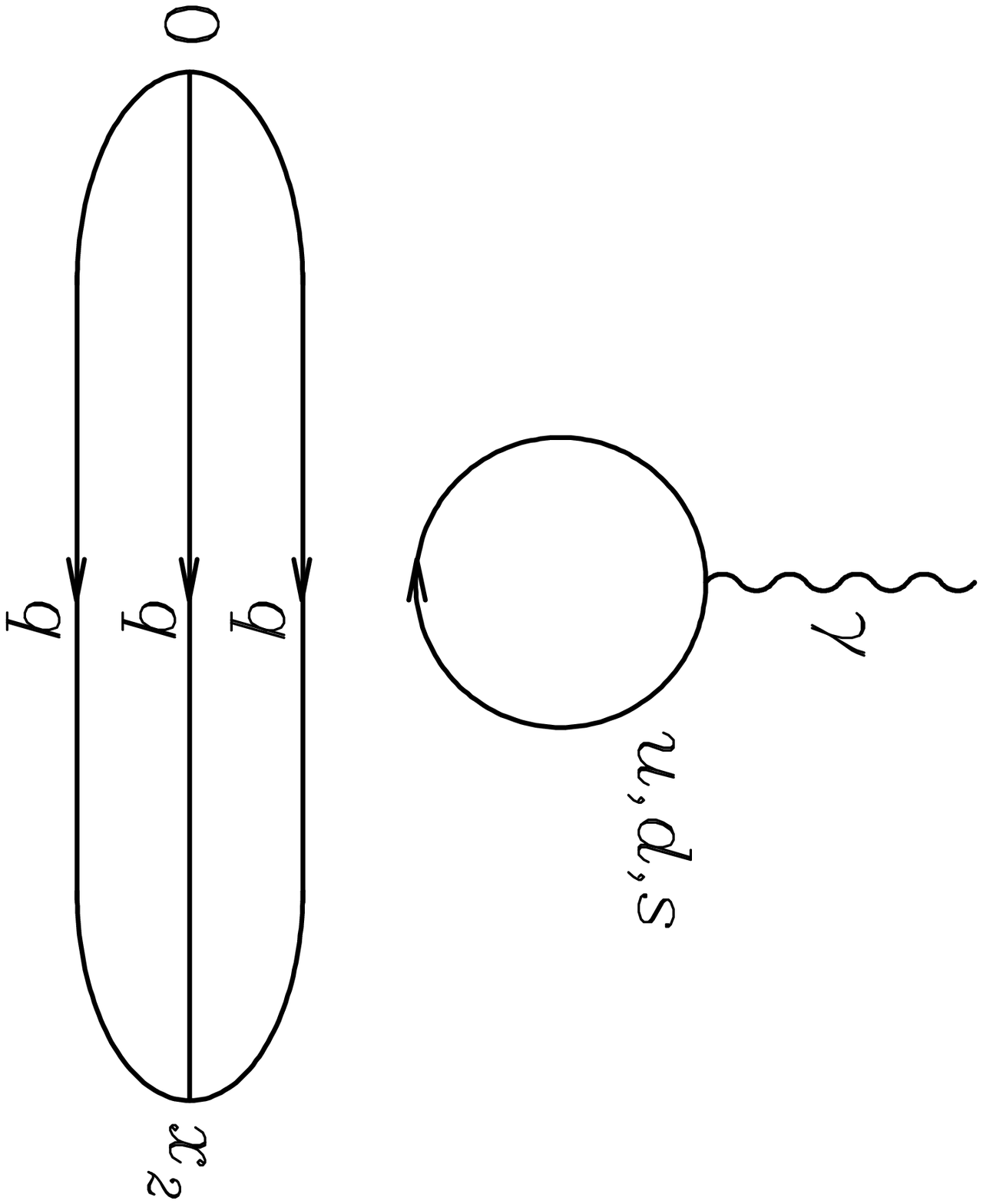,height=4cm}}}\end{picture}}
\end{picture}
\end{center}
\caption{Diagrams illustrating the two topologically different
insertions of the current within the framework of lattice QCD.  
These skeleton diagrams for the connected
(left) and disconnected (right) current insertions may be dressed by
an arbitrary number of gluons and quark loops.}
\label{topology}
\end{figure}

An examination of the symmetries manifest in the QCD path integral for
current matrix elements reveals various relationships among the quark
sector contributions \cite{Leinweber:1996ie}.  Current matrix elements
of hadrons, such as the magnetic moment of the proton, are extracted
from the three-point function, a time-ordered product of three
operators\cite{Leinweber:1996ie}.  Generally, an operator exciting the
hadron of interest from the QCD vacuum is followed by the current of
interest, which in turn is followed by an operator annihilating the
hadron back to the QCD vacuum.  In calculating the three point
function, one encounters two topologically distinct ways of
performing the electromagnetic current insertion.  Figure
\ref{topology} displays skeleton diagrams for these two possible
insertions (with Euclidean time increasing to the right).  These
diagrams may be dressed with an arbitrary number of gluons and quark
loops.  The left-hand diagram illustrates the connected insertion of
the current to one of the ``valence''%
\footnote{Note that the term ``valence'' used here differs from that
commonly used in the framework of deep-inelastic scattering.
Here ``valence'' simply describes the quark whose quark
flow line runs continuously from $0 \to x_2$.  These lines can flow
backwards as well as forwards in time and therefore have a sea
contribution associated with them \protect\cite{dblPiCloud}.}
quarks of the baryon.  In the right-hand diagram the external field
produces a $q \, \overline q$ pair which in turn interacts with the
valence quarks of the baryon via gluons. It is important to realize
that within the lattice QCD calculation of the loop diagram on the
right in Fig.\ \ref{topology} there is no antisymmetrization (Pauli
blocking) of the quark in the loop with the valence quarks. For this
reason, in general {\em only the sum of the two processes in Fig.\
\ref{topology} is physical}.

\section{CHARGE SYMMETRY EQUALITIES}

Under the assumption of charge symmetry, the three-point correlation
functions for octet baryons leads to the following equalities for
electromagnetic current matrix elements \cite{Leinweber:1996ie}:
\begin{eqnarray}
p =& e_u\, u_p + e_d\, d_p + O_N  \, , \qquad
n &= e_d\, u_p + e_u\, d_p + O_N  \, ,  \nonumber \\
\Sigma^+ =& e_u\, u_{\Sigma^+} + e_s\, s_\Sigma + O_\Sigma  \, , \qquad
\Sigma^- &= e_d\, u_{\Sigma^+} + e_s\, s_\Sigma + O_\Sigma  \, , \nonumber  \\
\Xi^0 =& e_s\, s_\Xi + e_u\, u_{\Xi^0} + O_\Xi  \, ,  \qquad
\Xi^- &= e_s\, s_\Xi + e_d\, u_{\Xi^0} + O_\Xi  \, .
\label{equalities}
\end{eqnarray}
Here, $O$ denotes the contributions from the quark-loop sector --
shown on the right-hand side of Fig. \ref{topology}.  The baryon label
represents the magnetic moment.  Subscripts allow for environment
sensitivity implicit in the three-point function
\cite{Leinweber:1996ie}.  For example, the three-point function for
$\Sigma^+$ is the same as for the proton, but with $d$ replaced by
$s$.  Hence, the $u$-quark propagators in the $\Sigma^+$ are
multiplied by an $s$-quark propagator, whereas in the proton the
$u$-quark propagators are multiplied by a $d$-quark propagator.  The
different mass of the neighboring quark gives rise to an environment
sensitivity in the $u$-quark contributions to observables, which means
that the naive expectations of the constituent quark model
$u_p/u_{\Sigma^+} = u_n/u_{\Xi^0} = 1$ may not be satisfied
\cite{Leinweber:1996ie,dblOctet,dblMagMomSR,dblDecuplet,%
dblDiquarks,dblShedLight,dblEssential}.  This observation should be
contrasted with the common assumption that the quark magnetic moment
is an intrinsic-quark property which is independent of the quark's
environment.

Focusing now on the nucleon, we note that for magnetic properties,
$O_N$ contains sea-quark-loop contributions from primarily $u$, $d$,
and $s$ quarks.  In the SU(3)-flavor limit ($m_u = m_d = m_s$) the
charges add to zero and hence the sum vanishes.  However, the heavier
strange quark mass allows for a result which is non-zero.  By
definition
\begin{eqnarray}
O_N &=& {2 \over 3} \,{}^{\ell}G_M^u - {1 \over 3} \,{}^{\ell}G_M^d -
{1 \over 3} \,{}^{\ell}G_M^s \, , \\
&=& {\,{}^{\ell}G_M^s \over 3} \left ( {1 - \,{}^{\ell}R_d^s \over
\,{}^{\ell}R_d^s } \right ) \, , \quad \mbox{where} \quad
{}^{\ell}R_d^s \equiv {\,{}^{\ell}G_M^s \over \,{}^{\ell}G_M^d} \, ,
\label{OGMs}
\end{eqnarray}
and the leading superscript, $\ell$, reminds the reader that the
contributions are loop contributions.  Note that, in deriving
Eq.(\ref{OGMs}), we have set ${}^{\ell}G_M^u = {}^{\ell}G_M^d$,
corresponding to $m_u = m_d$ \cite{Leinweber:1996ie}.  In the
constituent quark model $\,{}^{\ell}R_d^s = m_d/m_s \simeq 0.65$.

With no more than a little accounting, the strange-quark loop
contributions to the nucleon magnetic moment, $G_M^s$ may be isolated
from (\ref{equalities}) and (\ref{OGMs}) in the following two
phenomenologically useful forms,
\begin{equation}
G_M^s = \left ( {\,{}^{\ell}R_d^s \over 1 - \,{}^{\ell}R_d^s } \right ) \left [
2 p + n - {u_p \over u_{\Sigma^+}} \left ( \Sigma^+ - \Sigma^- \right
) \right ] \, ,
\end{equation}
and
\begin{equation}
G_M^s = \left ( {\,{}^{\ell}R_d^s \over 1 - \,{}^{\ell}R_d^s } \right ) \left [
p + 2n - {u_n \over u_{\Xi^0}} \left ( \Xi^0 - \Xi^- \right ) 
 \right ] \, .
\end{equation}
As we explained above, under the assumption that quark magnetic
moments are not environment dependent, these ratios (i.e.
${u_p}/{u_{\Sigma^+}}$ and ${u_n}/{u_{\Xi^0}}$) are taken to be one in
many quark models.  Incorporating the experimentally measured baryon
moments leads to:
\begin{equation}
G_M^s = \left ( {\,{}^{\ell}R_d^s \over 1 - \,{}^{\ell}R_d^s } \right ) \left [
3.673 - {u_p \over u_{\Sigma^+}} \left ( 3.618 \right ) \right ] \, , 
\label{ok}
\end{equation}
and
\begin{equation}
G_M^s = \left ( {\,{}^{\ell}R_d^s \over 1 - \,{}^{\ell}R_d^s } \right ) \left [
-1.033 - {u_n \over u_{\Xi^0}} \left ( -0.599 \right ) \right ] \, ,
\label{great}
\end{equation}
where all moments are expressed in nuclear magnetons $(\mu_N)$. (Note
that the measured magnetic moments are all known sufficiently
accurately\cite{PDG} that the experimental errors play no role in our
subsequent analysis.)  We stress that {\em these expressions for
$G_M^s$ are exact consequences of QCD, under the assumption of charge
symmetry}. Equation (\ref{great}) provides a particularly favorable
case for the determination of $G_M^s$ with minimal dependence on the
valence-quark ratio.

If one considers the quark model suggestions of $u_n/u_{\Xi^0}=1$ and
${}^{\ell}R_d^s=0.65$ in (\ref{great}), one finds $G_M^s = -0.81\
\mu_N$, a significant departure from the experimental indication of
positive values.

\begin{figure}[tbp]
\centering{\
\rotate{\epsfig{file=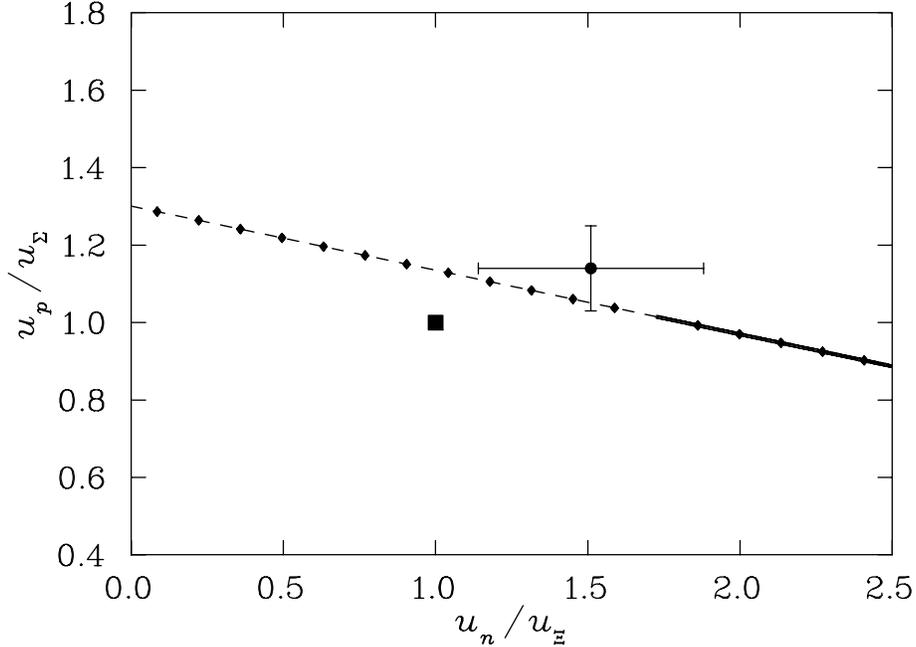,height=12cm}} }
\caption{The consistency relation between $u_p/u_{\Sigma^+}$ and
$u_n/u_{\Xi^0}$ which must be satisfied within QCD.  The part of the
straight line which is dashed corresponds to $G_M^s(0) < 0$, while the
solid part of the line has $G_M^s(0) > 0$.  The standard quark model
assumption of intrinsic quark moments independent of their environment
is indicated by the filled square at $(1, 1)$.  The diamonds on the
constraint line mark 0.1 $\mu_N$ changes in $G_M^s(0)$ when
${}^{\ell}R_d^s = 0.55$ \protect\cite{liuDisconn}.  The lattice QCD
prediction (after an appropriate chiral extrapolation, discussed in
following sections) is illustrated by the filled circle with standard
errors indicated.  }
\label{SelfCons}
\end{figure}

Equating (\ref{ok}) and (\ref{great}) provides a linear relationship
between $u_p/u_{\Sigma^+}$ and $u_n/u_{\Xi^0}$ which must be satisfied
within QCD. Figure \ref{SelfCons} displays this relationship by the
dashed and solid line, the latter corresponding to values for which
$G_M^s(0) > 0$ when ${}^{\ell}R_d^s$ is in the anticipated range $0 <
{}^{\ell}R_d^s < 1$.  Since the line does not pass through the point
$(1, 1)$ corresponding to the simple quark model assumption of
universality, the experimentally measured baryon moments are signaling
that there must be an environment effect exceeding 12\% in both ratios
or approaching 20\% or more in at least one of the ratios.  Moreover,
a positive value for $G_M^s(0)$ requires an environment sensitivity
exceeding 70\% in the $u_n/u_{\Xi^0}$ ratio.  Hence the experimental
indication that $G_M^s(0) > 0$ challenges the intrinsic magnetic
moment concept which is fundamental to the constituent quark model.

\section{CHIRAL CORRECTIONS}
\label{sec:chiral}
 
\begin{figure}[tb]
\centering{\
\rotate{\epsfig{file=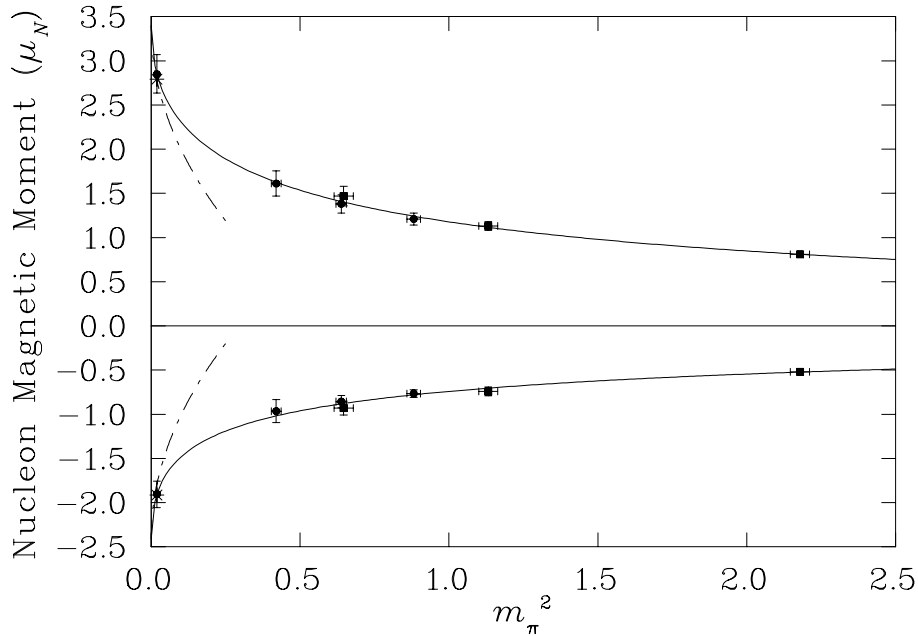,height=12cm}} }
\caption{Extrapolation of lattice QCD magnetic moments ($\bullet$
Ref.\ \protect\cite{dblOctet} and $\scriptstyle\blacksquare$  Ref.\
\protect\cite{wilcox92}) for the proton (upper) and neutron (lower) to
the chiral limit.  The solid curves illustrate a two parameter fit of
Eq.~(\protect\ref{mag-mom}) to the simulation data in which the
one-loop corrected chiral coefficient of $m_\pi$ is taken from
$\chi$PT.  The dashed curves illustrate the leading non-analytic
contribution of Eq.~(\protect\ref{LNAeq}).  The experimentally
measured moments are indicated by asterisks.  }
\label{LNAcomp} 
\end{figure}

One of the major challenges at present in connecting lattice
calculations of hadronic properties with the physical world is that
computational limitations restrict the accessible quark masses to
values an order of magnitude larger than the physical values. At such
large masses one is far from the region where chiral perturbation
theory is applicable.  Yet one knows that for current quark masses
near zero there is important non-analytic structure (as a function of
the quark mass) which must be treated correctly if we are to compare
with physical hadron properties. Our present analysis of the
strangeness magnetic form factor has been made possible by a recent
breakthrough in the treatment of these chiral corrections for the
nucleon magnetic moments \cite{Leinweber:1998ej,Thomas:1999mv}. In
particular, a study of the dependence of the nucleon magnetic moments
on the input current quark mass, within a chiral quark model which was
fitted to existing lattice data, suggested a model independent method
for extrapolating baryon magnetic moments which satisfied the chiral
constraints imposed by QCD.  We briefly summarize the main results of
that analysis:
\begin{itemize}
\item a series expansion of $\mu _{p(n)}$ in powers of $m_{\pi }$ is
not a valid approximation for $m_{\pi }$ larger than the physical
mass,
\item on the other hand, the behavior of the model, after adjustments
to fit the lattice data at large $m_{\pi }$ is well determined by
the simple Pad\'e approximant:
\begin{equation}
\mu _{p(n)}=\frac{\mu _{0}}{1-\frac{\chi }{\mu _{0}}m_{\pi }+ c\,
m^{2}_{\pi }} \, ,
\label{mag-mom}
\end{equation}
as illustrated in Fig.\ \ref{LNAcomp}.

\item Eq.(\ref{mag-mom}) not only builds in the usual magnetic moment 
of a Dirac particle at
moderately large $m^{2}_{\pi }$ but has the correct leading non-analytic
(LNA) behavior of chiral perturbation theory
\begin{equation}
\mu =\mu _{0} + \chi\, m_{\pi} ,
\label{LNAeq}
\end{equation}
with $\chi$ a model independent constant.
\item fixing $\chi$ at the value given by chiral perturbation theory
and adjusting $\mu _{0}$ and $c$ to fit the lattice data yielded
values of $\mu _{p}$ and $\mu _{n}$ of $2.85\pm 0.22\, \mu _{N}$ and
$-1.96\pm 0.16\, \mu _{N}$, respectively, at the physical pion
mass. These are in remarkably good agreement with the experimental
values -- certainly much closer than the usual linear extrapolations
in $m_{q}$.
\end{itemize}

Clearly it is vital to extend the lattice calculations of the baryon
magnetic moments to lower values of $m_{\pi }$ than the 600 MeV used
in the study just outlined.  It is also vital to include dynamical
quarks.  Nevertheless, the apparent success of the extrapolation
procedure gives us strong encouragement to investigate the same
approach for resolving the strange quark contribution to the proton
magnetic moment.

Our focus here is to extrapolate existing lattice QCD estimates of
valance quark contributions to baryon moments $(u_p,\ u_n,\
u_{\Sigma^+},\ u_{\Xi^0})$ to the physical mass regime.  The leading
non-analytic pieces of these various contributions to the baryon
magnetic moments are proportional to
\begin{equation}
\beta \, {m_N \over 8 \, \pi \, f_\pi^2} \, m_\pi \equiv \chi \, m_\pi \, ,
\label{LNAdef}
\end{equation}
where the pion decay constant $f_\pi = 93$ MeV.  Isolating a
particular quark flavour contribution only requires setting the
electric charge of all other quark flavours to zero.  Separation of
the contributions of the LNA terms into valance and sea-quark sectors
has been resolved in Ref.\ \cite{Leinweber:1999nf}.  The coefficients
$\beta$ and $\chi$ for various quark sectors and baryons are
summarized in Table \ref{table:chiralCoeff}.  In removing the $u$
sea-quark contribution from the $u$-quark sector, one eliminates the
contributions of the $\overline u$ sea-quark to the LNA term.  Since
the $z$-component of angular momentum in the pion-baryon system is
positive, the $\overline u$ contribution to the moment is negative.
Hence its removal suppresses the non-analyticity in the neutron moment
and enhances the non-analyticity in $\Xi^0,\ p,$ and $\Sigma^+$ where
the pion cloud acts to enhance the $u$-quark sector contribution to
these baryon moments.

\begin{table}[tbp]
\caption{Coefficients $\beta$ and $\chi$ of (\protect\ref{mag-mom})
and (\protect\ref{LNAdef}) for the leading nonanalytic (LNA)
contribution of the $u$-quark(s) to the magnetic moment of various
baryons. We express the LNA coefficients $\chi$ in terms of the usual
SU(6) constants $F$ and $D$, as well as giving numerical values for
$\chi$ and $\beta$ (which are equivalent through
Eq.(\protect\ref{LNAdef})). The $u$-quark charge has been normalized
to 1.}
\label{table:chiralCoeff}
\begin{tabular}{lcccc}
\hline
Baryon &$\beta, \chi$ &$u$-flavor sector &$u$-sea-quark loop
&$u$-valence sector \\ 
\hline 
$n$        &$\beta$     
           &$(F+D)^2$ 
           &$(9\, F^2 - 6\, FD + 5\, D^2)/3$
           &$2 (-3\, F^2 + 6\, FD - D^2)/3$ \\
           &$\beta$  &$1.020$ &$0.612$ &$0.408$ \\ 
           &$\chi$   &$4.41$  &$2.65$  &$1.76$  \\
$\Xi^0$    &$\beta$     
           &$-(F-D)^2$ 
           &$+(F-D)^2$ 
           &$-2(F-D)^2$ \\
           &$\beta$  &$-0.0441$ &$0.0441$ &$-0.0882$ \\ 
           &$\chi$   &$-0.191$  &$0.191$  &$-0.381$  \\
$p$        &$\beta$     
           &$-(F+D)^2$ 
           &$(9\, F^2 - 6\, FD + 5\, D^2)/3$
           &$-4 (3\, F^2 + 2\, D^2)/3$ \\
           &$\beta$  &$-1.020$ &$0.612$ &$-1.632$ \\ 
           &$\chi$   &$-4.41$  &$2.65$  &$-7.06$  \\
$\Sigma^+$ &$\beta$     
           &$-2 (3\, F^2 + D^2)/3$
           &$2 (3\, F^2 + D^2)/3$
           &$-4 (3\, F^2 + D^2)/3$ \\
           &$\beta$  &$-0.568$ &$0.568$ &$-1.136$ \\ 
           &$\chi$   &$-2.46$  &$2.46$  &$-4.91$ \\
\hline
\end{tabular}
\end{table}

For the calculation of $G_M^s(0)$, the LNA chiral behavior of $u_n$
and $u_{\Xi^0}$ is crucial, as the corresponding coefficients are of
opposite sign.  As a result, the ratio of these $u$-quark
contributions, extrapolated with the correct chiral behavior, will be
completely different from the linearly extrapolated ratio.  That the
signs of these chiral coefficients are opposite may be traced back to
the charge of the predominant pion cloud associated with these
baryons.  The chiral behavior in the neutron case is dominated by the
transition $n \to p\, \pi^- \to n$, where there is a contribution from
a $\bar{u}$ loop.  On the other hand, for the $\Xi^0$ the LNA
$u$-quark contribution is dominated by the process $\Xi^0 \to \Xi^-
\pi^+ \to \Xi^0$.  In this case the virtual $\pi^+$ involves a valence
$u$-quark -- and hence the sign change.

In order to extrapolate the lattice data for valence quarks to the
physical pion mass we use the same approach which worked so well for
the neutron and proton moments in Ref.\
\cite{Leinweber:1998ej,Thomas:1999mv}.  That is, we fit the lattice
data for the valence quark of flavor $q$ in baryon $B$ with the form:
\begin{equation}
\mu(m_\pi) = \frac{\mu^{0}}
{1 - \frac{\chi}{\mu^0} \, m_\pi + c \, m_\pi^2}.
\label{eq:chiral}
\end{equation}
Here $\chi$ are the modified LNA chiral coefficients for the valence
quarks given in Table \ref{table:chiralCoeff} (last column) and
$\mu^0$ and $c$ are fitting parameters.

\begin{figure}[tb]
\centering{\
\rotate{\epsfig{file=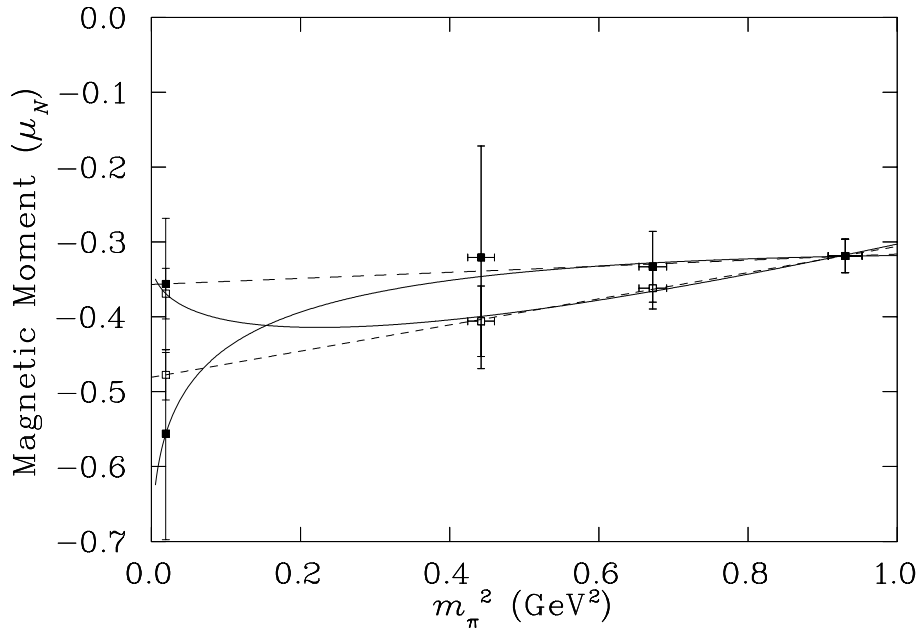,height=12cm}} }
\caption{Extrapolation of the valence $u$-quark magnetic moment
contributions to the neutron (solid symbols) and $\Xi^0$ (open
symbols).  Naive linear extrapolations (dashed lines) are contrasted
with the chiral extrapolations incorporating the leading nonanalytic
behavior.  The lattice data is taken from Ref.\
\protect\cite{dblOctet}.  }
\label{NXiExtrap}
\end{figure}

\begin{figure}[tb]
\centering{\
\rotate{\epsfig{file=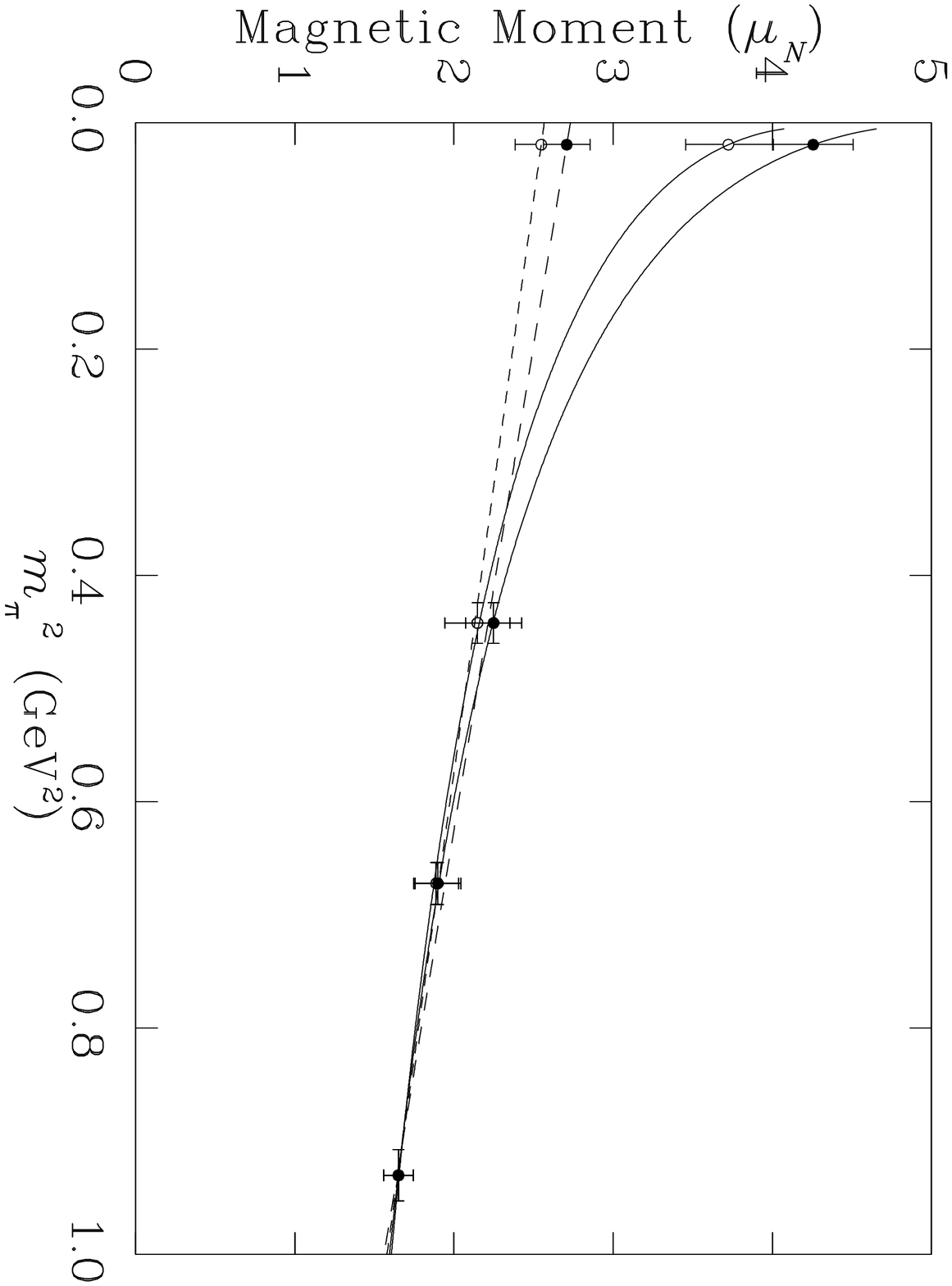,height=12cm}} }
\caption{Extrapolation of the valence $u$-quark magnetic moment
contributions to the proton (solid symbols) and $\Sigma^+$ (open
symbols).  Naive linear extrapolations (dashed lines) are contrasted
with the chiral extrapolations based on (\protect\ref{eq:chiral}).
The lattice data is taken from 
Ref.\ \protect\cite{dblOctet}.
}
\label{PSigExtrap}
\end{figure}

Figures \ref{NXiExtrap} and \ref{PSigExtrap} illustrate the
extrapolations of the lattice data.  The results of these fits shown
at $m_\pi = 140$ MeV are the results of the extrapolation procedure,
including fitting errors.  Naive linear extrapolations are also shown
to help emphasize the dominant role of chiral symmetry in the
extrapolation process.

Clearly the effect of the chiral corrections on the extrapolated
values of the key magnetic moment ratio $u_n / u_{\Xi^0}$ at
$m_\pi=140$ MeV is dramatic.  Instead of the value $0.72 \pm 0.46$
obtained by linear extrapolation one now finds $1.51 \pm 0.37 $.  It
is vital that this ratio is much more consistent with the range of
values found necessary to yield a positive value of $G_M^s$, as
illustrated in Fig.\ \ref{SelfCons} -- values which seemed quite
unreasonable in the constituent quark model.  By comparison, the ratio
$u_p / u_{\Sigma^+}$ at $m_\pi = 140$ MeV is quite stable, changing
from $1.14 \pm 0.08\ $ in the case of linear extrapolation to $1.14
\pm 0.11 $ when the correct chiral extrapolation is used.  The lattice
point illustrated in Fig.\ \ref{SelfCons} is consistent with the
constraint obtained from charge symmetry.  This suggests that any
charge symmetry breaking in the experimentally measured moments is
small.

The effect of the chiral extrapolation on the ratio
${u_n}/{u_{\Xi^0}}$ is the most dramatic, changing it from $0.72 \pm
0.46$ to $1.51 \pm 0.37$. Thus Eq. (\ref{great}), which seemed certain
to guarantee $G_M^s < 0$, no longer does.  Indeed, using the recent
estimate \cite{liuDisconn} of ${}^{\ell}R_d^s = 0.55$, we obtain
$G^s_M = -0.16 \pm 0.18 \mu_N$.  While this is still negative it does
permit a small positive value within the error.

\end{document}